\documentclass[%
 aip,
 amsmath,amssymb,
 reprint,%
]{revtex4-1}

\usepackage{graphicx}
\usepackage{dcolumn}
\usepackage{bm}
\usepackage{ulem}
\usepackage[utf8]{inputenc}
\usepackage[T1]{fontenc}
\usepackage{mathptmx}
\usepackage{subfigure}
\usepackage{etoolbox}
\usepackage{dsfont}
\usepackage{amsthm}
\usepackage[ruled,linesnumbered]{algorithm2e}
\newtheorem{theorem}{Theorem}
\usepackage{mathtools}
\usepackage{tikz}
\usepackage{float}

\newcommand*\circled[1]{\tikz[baseline=(char.base)]{
            \node[shape=circle,draw,inner sep=0.8pt] (char) {#1};}}

\newcommand{\rv}[1]{{\color{blue}#1}}

\newcommand{\ket}[1]{\lvert#1\rangle} 

\newcommand{\braopket}[3]{\langle #1 | #2 | #3\rangle} 

\RestyleAlgo{ruled}

\makeatletter
\def\@email#1#2{%
 \endgroup
 \patchcmd{\titleblock@produce}
  {\frontmatter@RRAPformat}
  {\frontmatter@RRAPformat{\produce@RRAP{*#1\href{mailto:#2}{#2}}}\frontmatter@RRAPformat}
  {}{}
}%
\makeatother
\begin{document}

\preprint{AIP/123-QED}

\title{QuGStep: Refining Step Size Selection in Gradient Estimation for Variational Quantum Algorithms}
\author{Senwei Liang}\thanks{These authors contributed equally and are listed alphabetically.}
  \affiliation{Lawrence Berkeley National Laboratory, Berkeley, CA 94720, U.S.A}
\author{Linghua Zhu}\thanks{These authors contributed equally and are listed alphabetically.}
\affiliation{ 
Department of Chemistry, University of Washington, Seattle, WA 98195, U.S.A
}
\author{Xiaosong Li}\thanks{Correspondence can be addressed to \textit{xsli@uw.edu} and \textit{cyang@lbl.gov}}
\affiliation{ 
Department of Chemistry, University of Washington, Seattle, WA 98195, U.S.A
}
\author{Chao Yang}\thanks{Correspondence can be addressed to \textit{xsli@uw.edu} and \textit{cyang@lbl.gov}}
    \affiliation{Lawrence Berkeley National Laboratory, Berkeley, CA 94720, U.S.A}

\date{\today}

\begin{abstract}
Variational quantum algorithms (VQAs) offer a promising approach to solving computationally demanding problems by combining parameterized quantum circuits with classical optimization. Estimating probabilistic outcomes on quantum hardware requires repeated measurements (shots). However, in practice, the limited shot budget introduces significant noise in the evaluation of the objective function. Gradient estimation in VQAs often relies on the finite-difference, which evaluates the noisy objective function at perturbed circuit parameter values. The accuracy of this estimation is highly dependent on the choice of step size for these perturbations. An inappropriate step size can exacerbate the impact of noise, causing inaccurate gradient estimates and hindering the classical optimization in VQAs. This paper proposes QuGStep, an algorithm that addresses the challenge of determining the appropriate step size for finite-difference gradient estimation under a shot budget. QuGStep is grounded in a theorem that proves the optimal step size, which accounts for the shot budget, minimizes the error bound in gradient estimation using finite differences. Numerical experiments approximating the ground state energy of several molecules demonstrate that QuGStep can identify the appropriate step size for the given shot budget to obtain effective gradient estimation. Notably, the step size identified by QuGStep achieved convergence to the ground state energy with over 94\% fewer shots compared to using a default step size  (\textit{i.e.,} step size of $0.01$). These findings highlight the potential of QuGStep to improve the practical deployment and scalability of quantum computing technologies.

\end{abstract}

\maketitle

\section{Introduction}
Quantum computing marks a significant leap beyond traditional computing methods by harnessing the principles of quantum mechanics~\cite{Cao19_10856, Meglio24_037001}. These principles enable quantum computers to tackle problems that classical computers struggle to handle efficiently. Examples include factoring large numbers~\cite{Shor94_124,Shor99_33}, simulating intricate molecular systems for pharmaceutical and materials research~\cite{Kandala17_7671}, and solving large linear equations to enhance data analysis capabilities~\cite{Subacsi19_060504} and so on. In addition, integrating quantum algorithms with artificial intelligence~\cite{Cai15_110504,Mangini21_10002,Dunjko18_074001} shows promising advancements in processing speeds and decision-making accuracy.

\begin{figure*}[ht]
\centering
{\includegraphics[width=0.95\linewidth ]{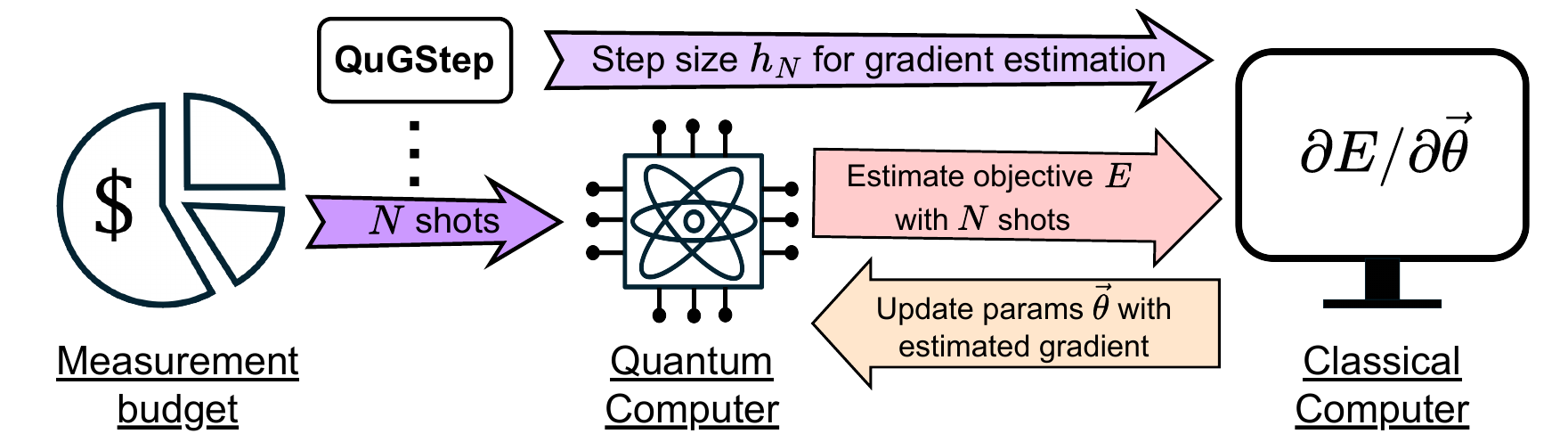}}
\caption{Semantic illustration of the relationship between the proposed QuGStep and VQA. In VQA, the quantum computer's outcome is evaluated by performing a specified number of measurements (shots), while the classical computer optimizes the quantum circuit parameters $\vec\theta$. Given the specified shot budget, QuGStep determines an appropriate step size for finite-difference gradient estimation.}
\label{fig:1}
\end{figure*}

Variational Quantum Algorithms (VQAs) are one of the most effective strategies in this emerging computing paradigm~\cite{Cerezo21_625}. These algorithms are designed for the current generation of quantum devices, often called near-term quantum devices. The basic idea behind VQAs is to use a parameterized quantum circuit to prepare a quantum state that encodes the solution to an optimization problem. The parameters of the quantum circuit are then classically optimized to minimize the objective function. This process, including the interaction between quantum and classical computers, is illustrated in Figure~\ref{fig:1}. This hybrid approach is capable of tackling complex challenges across various fields, including quantum chemistry, finance, and combinatorial optimization. In quantum chemistry, in particular, the Variational Quantum Eigensolver (VQE)~\cite{McClean16_023023} is a specific implementation of a VQA tailored for solving quantum chemistry problems. The VQE aims to find the ground-state energy of a molecule and the corresponding wavefunction of the molecular system.

Quantum computations inherently produce probabilistic outcomes. To evaluate the objective function, repeated measurements are required, where a single measurement is termed ``one shot''~\cite{Holevo11_book}. The measurement result is averaged to reduce the effect of \textit{shot noise}. The reduction of shot noise follows the statistical error scaling law, specifically $\mathcal{O}(1/\sqrt{N})$, where $N$ is the number of shots. Although increasing the number of shots improves the accuracy of the objective function evaluation, it is costly for each shot~\cite{Phalak23_41514}.


In VQA, gradient-based classical optimization algorithms are often used, which necessitates the computation of the gradient of the objective function with respect to the circuit parameters. Several popular gradient evaluation methods are available for this purpose, including the parameter shift rule~\cite{Crooks19_arXiv,Wierichs22_677}, finite difference~\cite{Guerreschi17_arXiv,Kan24_arXiv}, natural gradient~\cite{Yamamoto19_arXiv,Stokes20_269,Wierichs20_043246,Koczor22_062416}, and machine-learning-based surrogate models~\cite{Luo24_36}. The parameter shift rule offers precise gradient calculations by systematically perturbing the circuit parameters with a fixed shift and observing the changes in the output~\cite{Warren22_arXiv}. However, it requires equidistant eigenvalues of the gate generators and fails when gates have non-uniform or more than two unique eigenvalues, thus limiting its applicability in more complex quantum circuits. The parameter shift rule needs $2d+1$ objective function evaluations, where $d$ denotes the number of circuit parameters. Another method, the natural gradient, adapts to the geometry of the parameter space, potentially offering more efficient convergence. However, calculating natural gradients can be computationally intensive as it often involves inverting or manipulating large matrices, which may limit its practicality on current quantum hardware, particularly for large-scale systems. The natural gradient method typically requires around $d^2+d$ objective function evaluations. Furthermore, machine learning (ML)-based techniques like QuACK~\cite{Luo24_36}, which utilizes Koopman operator learning, can forecast parameter updates, reducing the necessary number of gradient estimations. However, ML-based methods may require sufficient gradient estimation as training data, and training an effective ML model typically necessitates some hyperparameter tuning. Lastly, finite-difference methods offer a straightforward approach requiring only $d+1$ function evaluations; however, they are susceptible to noise~\cite{Uvarov20_075104,Mari21_012405,Lai23_11,Teo24_012620}. Each of these methods has its own advantages and is better suited for specific quantum computational tasks. Note that finite-difference methods remain essential in several practical scenarios. Not all parameterized quantum gates satisfy the eigenvalue requirements for the parameter shift rule, particularly gates with continuous parameters or those generated by non-Pauli operators~\cite{Wierichs22_677}. Furthermore, pulse-level quantum control, which achieves state-of-the-art gate fidelities, naturally aligns with finite-difference approaches due to hardware discretization and model uncertainties~\cite{Werninghaus21_14}. This paper focuses on the finite difference method due to its straightforward and generic approach to gradient estimation.

When estimating the gradient using the finite difference method in VQA, the noisy objective function should be evaluated at parameters perturbed by a small step size $h$ and the accuracy of this estimation is highly dependent on the choice of $h$. 
As we will discuss in Section~\ref{sec:error}, the total error in finite-difference gradient estimation comprises two parts: the inherent truncation error, which is $\mathcal{O}(h)$, and the error from shot noise, which is $\mathcal{O}(1/(\sqrt{N}h))$, where $N$ denotes the number of shots. If $h$ is too small, the error from shot noise will dominate. Conversely, if $h$ is too large, the truncation error will be significant. This relationship underscores the importance of selecting an appropriate step size in VQAs, given the limited shot budget $N$. While existing studies recognize this issue, they rely on empirical selection and lack a clear strategy or theoretical framework to address it~\cite{Harwood22_1,Kan24_arXiv}. As we shall see in Section \ref{sec:numerical}, an inappropriately small step size led to poor performance and required a significant increase in the number of shots to achieve a better convergence. Many well-developed software packages use a very small default $h$ for gradient estimation, such as $h=1.5\times 10^{-8}$ in the Python package \texttt{scipy.optimize.approx\_fprime}.~\footnote{https://docs.scipy.org/doc/scipy/reference/generated/\\scipy.sparse.linalg.expm\_multiply.html} Similarly, some empirical small step sizes, such as $h=0.01$, have been used in VQA~\cite{Guerreschi17_arXiv,Zhou20_021067}. However, these small step sizes are not well-suited or directly applicable to VQA without considering $N$. 

This paper proposes QuGStep, an algorithm that refines the step size $h$ for finit-difference gradient estimation under a shot budget, as shown in Figure~\ref{fig:1}.  QuGStep is grounded in Theorem~\ref{thm}, which proves the optimal step size (denoted as $h_N$) that minimizes the error bound in gradient estimation using finite differences with budget $N$. However, $h_N$ depends not only on $N$ but also on the second-order derivative estimates, making it impractical to use directly. In QuGStep, we first perform a few trial VQA runs using significantly fewer shots $\hat{N}$ to get an estimate of $h_{\hat{N}}$. We then determine $h_N$ using $h_{\hat{N}}$ based on the scaling relation with respect to the shot budget. 
In Section~\ref{sec:numerical}, by using the step size identified by QuGStep, we will see that the optimization process of VQE converges to ground state energies with significantly fewer shots—over 94\% fewer—compared to using a typical default step size (\textit{i.e.,} step size of $0.01$). We clarify that QuGStep is designed to identify the correct order of magnitude for the step size $h_N$ rather than pinpointing a single, uniquely optimal value, thereby avoiding an incorrect order of magnitude. The obtained $h_N$ should be regarded as a robust estimate within a range of well-performing values. The contribution of this paper not only provides mathematical guidance on choosing the step size but also highlights the potential of QuGStep to improve the practical deployment and scalability of quantum computing technologies. We should also note that while the proposed algorithm primarily focuses on shot noise, it may be extendable to other types of noise, such as hardware-specific gate-based noise (see Section~\ref{sec:discussion}).


The paper is organized as follows. Section \ref{sec:pre} introduces the procedure of VQA, with a focus on error analysis of the finite difference method for gradient estimation. Section \ref{sec:method} first proves the optimal step size for gradient estimation under shot noise, and then presents the proposed QuGStep for obtaining such a step size. The numerical experiments in Section \ref{sec:numerical} demonstrate the effectiveness of QuGStep. Finally, we discuss the findings of these experiments in Section \ref{sec:discussion}, and the paper concludes in Section \ref{sec:conclusion}.

\section{Preliminary}
\label{sec:pre}
\subsection{Variational Quantum Algorithms}
The goal of VQAs is to optimize a cost function associated with a quantum state by applying the variational principle. This hybrid quantum-classical approach involves iteratively updating parameters in a quantum circuit on a classical computer while evaluating an objective function using a quantum computer, as depicted in Figure~\ref{fig:1}.

Consider $U(\vec\theta)$ as a quantum circuit parameterized by $\vec\theta$. The parameterized quantum state can be represented as: $\ket{\psi(\vec\theta)} = U(\vec\theta) \ket{\psi_{\text{ref}}}$, where $\ket{\psi_{\text{ref}}}$ denotes a reference state chosen based on the specific problem. Different VQAs use different reference states: computational basis states for optimization problems or Hartree-Fock states for chemistry applications. The expectation value of an observable $\hat{O}$ can be computed as:
\begin{equation}
\begin{aligned}
  E(\vec\theta) = \braopket{\psi(\vec\theta)}{\hat{O}}{\psi(\vec\theta)}
  = \braopket{\psi_{\text{ref}}}{U^{\dagger}(\vec\theta) \hat{O} U(\vec\theta)}{\psi_{\text{ref}}}. 
\end{aligned}    
\label{eq:expected_value}
\end{equation}

The optimal solution is found by minimizing this expectation value through classical optimization techniques. VQAs have been applied to diverse applications including quantum chemistry, machine learning, optimization problems, and quantum simulation. In our approach, we apply this framework to molecule simulation with the Hamiltonian ($H$) as the observable $\hat{O}$, where $E(\vec\theta)$ corresponds to the energy. Minimizing $E(\vec\theta)$ determines the ground state energy of molecular systems.

\subsection{Objective Function and Gradient Evaluation}
\label{sec:error}
The objective function in Eqn.~\eqref{eq:expected_value} is evaluated on a quantum computer by taking repeat measurements.
Each measurement (shot) for $E(\vec\theta)$ can be viewed as drawing a random sample, $e^s(\theta)$, of a random variable with the mean $E(\vec\theta)$ and a standard deviation $\sigma(\vec\theta)$. Averaging multiple samples of $N$ shots, \textit{i.e.}, $e^s(\theta), s=1, 2, ..., N$, we have $\bar{E}(\vec\theta):=\frac{1}{N}\sum_{s=1}^Ne^s(\theta)$ to estimate $ E(\vec\theta)$. Therefore, $\bar{E}(\vec\theta)$ can be written as 
\begin{align}
    \bar{E}(\vec\theta)=E(\vec\theta)+\epsilon_N(\vec\theta),
    \label{eqn:estimator}
\end{align}
where $\epsilon_N(\vec\theta)$ represents shot noise, which can be considered a random variable with zero mean and variance $\frac{\sigma^2(\vec\theta)}{N}$. Based on the central limit theorem~\cite{Billingsley17_book}, the probability of $\bar{E}(\vec\theta)$ deviating from $E(\vec\theta)$ decreases as $N$ increases. 

In VQA, the optimization problem $\text{min}_{\vec\theta} \bar{E}(\vec\theta)$ is solved by updating $\vec\theta$ on a classical computer using a standard numerical optimization algorithm. Various optimizers, such as gradient descent, Adam~\cite{Kingma14_arXiv} and the Broyden-Fletcher-Goldfarb-Shanno method~\cite{Fletcher13_book} can be used to iteratively update the parameter $\vec\theta$:
\begin{align*}
    \vec\theta^{t+1} = \mathrm{Optimizer}(\vec\theta^t, \frac{\partial E(\vec\theta)}{\partial\vec\theta}|_{\vec\theta^{t}}, \gamma),
\end{align*}
where $\gamma$ denotes a learning rate and $t$ denotes iteration. When the optimizer is gradient descent, $\vec\theta^{t+1} = \vec\theta^{t}-\gamma\frac{\partial E(\vec\theta)}{\partial\vec\theta}|_{\vec\theta^{t}}$. 

Let $\vec\theta=(\theta_1,\cdots,\theta_d)$ with $d$ parameters. A finite difference method to estimate the gradient of the $i$th component gives:
\begin{align}
\frac{\partial E(\vec\theta)}{\partial \theta_i}&\approx \frac{\bar{E}(\vec\theta+he_i)-\bar{E}(\vec\theta)}{h},\quad i=1,\cdots,d,
\label{eqn:fd}
\\&=\underbrace{\frac{{E}(\vec\theta+he_i)-{E}(\vec\theta)}{h}}_{\circled{1}}+\underbrace{\frac{\epsilon_N(\vec\theta+he_i)-\epsilon_N(\vec\theta)}{h}}_{\circled{2}}
\label{eqn:grad}
\end{align}
where $h$ is the step size and $e_i$ is a one-hot vector with the $i$-th entry equal to 1 and all other entries equal to 0.
Such an approximation requires two evaluations, 
$E(\vec\theta)$ and $E(\vec\theta+he_i)$.

The total error in finite-difference gradient estimation~\eqref{eqn:grad} comprises two parts. The first part is the local truncation error (difference between \circled{1} and $\partial E(\vec\theta)/\partial \theta_i$), which is on the order of $\mathcal{O}(h)$\footnote{The notation $f(\alpha) = \mathcal{O}(\alpha)$ means: There exists a positive constant $C$ independent of $\alpha$ such that the absolute value of $f(\alpha)$ is at most $C\alpha$, i.e., $|f(\alpha)| \leq C\alpha$.} when $E(\vec\theta)$ is smooth~\cite{Iserles09_book}. The second part is the error from the shot noise. Informally, since $\epsilon_N(\vec\theta)$ has a mean of 0 and a standard deviation of $\sigma(\vec\theta)/\sqrt{N}$, we have $\epsilon_N(\vec\theta)=\mathcal{O}(\sigma(\vec\theta)/\sqrt{N})$, and hence the term $\circled{2}=\mathcal{O}(\sigma(\vec\theta)/(\sqrt{N}h))$. To balance these two competing errors, we seek an $h$ such that $\circled{2}= \mathcal{O}(h)$. Therefore, $\sqrt{N}h^2=\mathcal{O}(1)$, which implies $h=\mathcal{O}(1/N^{1/4})$. This establishes the relationship $N$ and $h$. In the next section, we will formally derive this relationship in the context of minimizing the error bound in  Eqn.~\eqref{eqn:grad}.

\section{Method}
\label{sec:method}
In this section, we formally introduce QuGStep and its theoretical foundation. First, in Section~\ref{sec:theory}, 
we present Theorem \ref{thm}, which demonstrates the optimal step size $h_N$ that minimizes the error bound of finite-difference gradient estimation given a shot budget $N$. Then, in Section~\ref{sec:qugstep}, we introduce QuGStep, a practical approach for determining $h_N$. 

\subsection{Optimal Step Size to Minimize Error Bound}
\label{sec:theory}
Without loss of generality, in this section, we consider the objective function $E({\theta})$ with a single variable  $\theta$. Given a fixed $\tilde{\theta}$, we seek an $h$ that minimizes the difference between the finite difference gradient estimation~\eqref{eqn:fd} and the true gradient  $E^{\prime}\left(\tilde{\theta}\right)$. This problem can be formulated as follows:
\begin{align}
\min _{h>0} \mathcal{E}(\tilde{\theta}, h):=\mathds{E}\left(\frac{\bar{E}\left(\tilde{\theta}+h\right)-\bar{E}\left(\tilde{\theta}\right)}{h}-E^{\prime}\left(\tilde{\theta}\right)\right)^2.
\label{obj}
\end{align}
 The expectation is taken with respect to the shot noise $\epsilon_N(\tilde{\theta})$ defined in Eqn.~\eqref{eqn:estimator}. Theorem~\ref{thm} provides an upper bound for $\mathcal{E}(\tilde{\theta},h)$ and determines the optimal $h$ (denoted as $h_N$) that minimizes this upper bound. The proof utilizes results from Ref.~\cite{More12_1} and is provided in the Appendix.

\begin{theorem} \label{thm} Let the function $E:\mathds{R}\to \mathds{R}$ be twice differentiable on the interval $I=(\tilde{\theta}-h_0,\tilde{\theta}+h_0)$ for some $h_0>0$ and let $\bar{E}(\theta)$ be defined in \eqref{eqn:estimator}. We assume that 

(i) the second-order derivative $|E''(\theta)|<\mu$, for $\theta\in I$;

    (ii) the standard deviation $\sigma(\theta)\leq \varsigma$ for $\theta\in I$.

\noindent Then, we have 

(1) $\mathcal{E}(\tilde{\theta}, h)$ has an upper bound: $\mathcal{E}(\tilde{\theta}, h)\leq \frac{1}{4}\mu^2h^2+\frac{2\varsigma^2}{h^2N}$;

(2) the upper bound is minimized when $h= h_N:=\frac{8^{1/4}\varsigma^{1/2}}{\mu^{1/2}N^{1 / 4}}$.

\end{theorem}


Although $h_N$ does not directly minimize $\mathcal{E}(\tilde{\theta}, h)$, it is a minimizer of the upper bound of $\mathcal{E}(\tilde{\theta}, h)$. This makes $h_N$ a reliable choice for the step size, as it effectively controls the error. 
In practice, obtaining $\mu$ and $\varsigma$ required to determine $h_N$ is impractical. This is particularly challenging because $\mu$ involves estimating the second-order derivative, which exceeds the original goal of finding the first-order derivative. Also, some estimates of $\mu$ might be too loose in practice (see Appendix~\ref{sec:errorbond}).
 
\subsection{QuGStep: A Practical Implementation}
\label{sec:qugstep}
In this section, we introduce QuGStep, which aims to find the step size $h_{N}$ for a given shot budget $N$. The basic idea behind QuGStep is to use a smaller test shot budget $\hat{N}$ to determine $h_{{N}}$ by performing VQA runs. We then scale $h_N$ to obtain $h_{\hat{N}}$ based on the scaling relation between $h_{N}$ and $h_{\hat{N}}$. Note that $\hat{N}$ is chosen to be much smaller than $N$ to minimize the overhead of determining $h_{\hat{N}}$. The workflow of QuGStep is summarized in Algorithm \ref{alg:grid}.

With Theorem~\ref{thm}, the scaling relation between $h_{N}$ and $h_{\hat{N}}$ with respect to the shot budgets $N$ and $\hat{N}$ can be expressed by:
\begin{align}
h_{N}=\frac{h_{\hat{N}}}{(N/\hat{N})^{1/4}}
\label{eqn:scale}
\end{align}
In practice, we typically use a constant step size throughout the optimization process, even though the definition of $h_N$ in Theorem~\ref{thm} is $\vec\theta$ dependent. Therefore, we consider both $h_{{N}}$ and $h_{\hat{N}}$ to be constant. Once we obtain $h_{\hat{N}}$, we can readily determine $h_{N_T}$ using Eqn.~\eqref{eqn:scale}.


To determine $h_{\hat{N}}$, we can perform a grid search. Specifically, we first set up a candidate set $\mathds{S}$. This set can be chosen on a logarithmic scale, for example, $\{\cdots, 10^{0}, 10^{-1}, 10^{-2}, \cdots\}$. When the candidate set is not large, we simply need to traverse all candidate values of $h$ in $\mathds{S}$. For each candidate value $h$, we perform a VQA with the given $N$, and then select the $h$ that yields the best performance profile as $h_{\hat{N}}$. To quantify the performance profile, we consider the average of the objective function values over the last 20 iterations.





\begin{algorithm}[hbt!]
\caption{QuGStep}\label{alg:grid}
\textbf{Input:} {Candidate set $\mathds{S}$, Target shot budget $N$, Test shot budget $\hat{N}$}\;
\textbf{Output:} Step size $h_{N}$ for $N$\;
\For{$h$ in $\mathds{S}$}{
    Perform VQE with $\hat{N}$ shots per objective function evaluation and step size $h$ for finite-difference gradient estimation\;
}
Select $h$ as $h_{\hat{N}}$ with the best performance profile, e.g., average of the objective function values over the last 20 iterations\;
Compute $h_{N}$ using Eqn.~\eqref{eqn:scale}.
\end{algorithm}


\begin{figure*}[ht]
\centering
	\subfigure[Test budget $\hat{N}=9$]{\includegraphics[width=0.33\linewidth ]{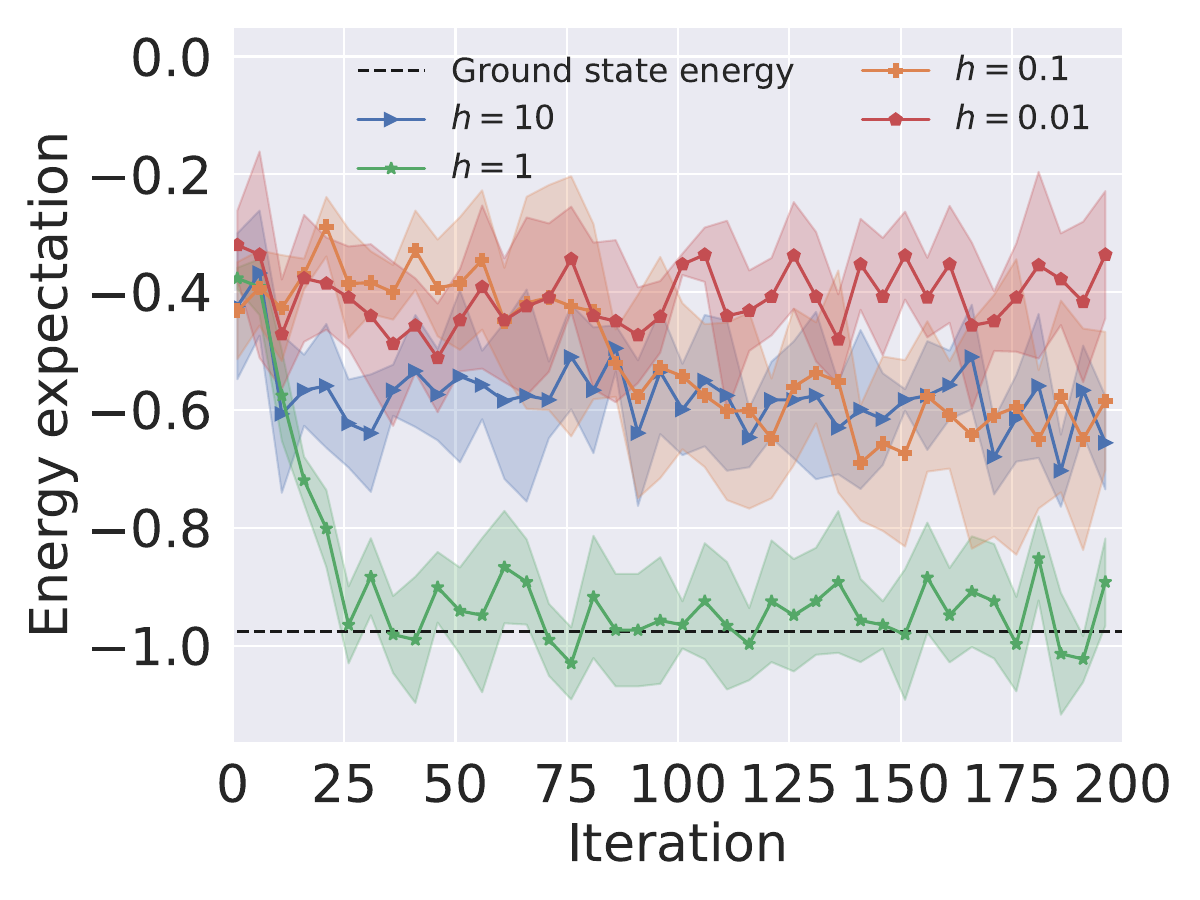}}
	\subfigure[Target budget $N=360$]{\label{H2_N360}\includegraphics[width=0.33\linewidth ]{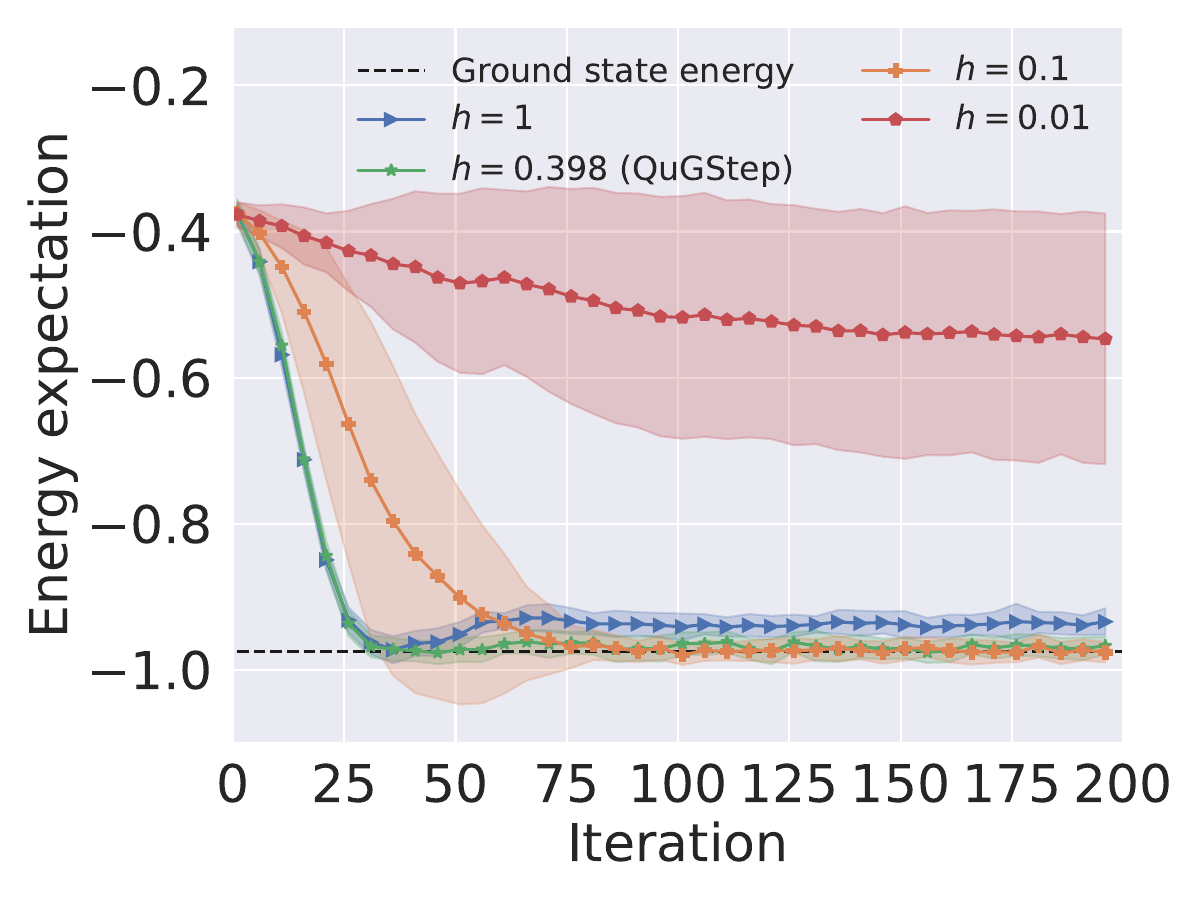}}
 \subfigure[{\color{blue}Varied $N$}]{\includegraphics[width=0.33\linewidth ]{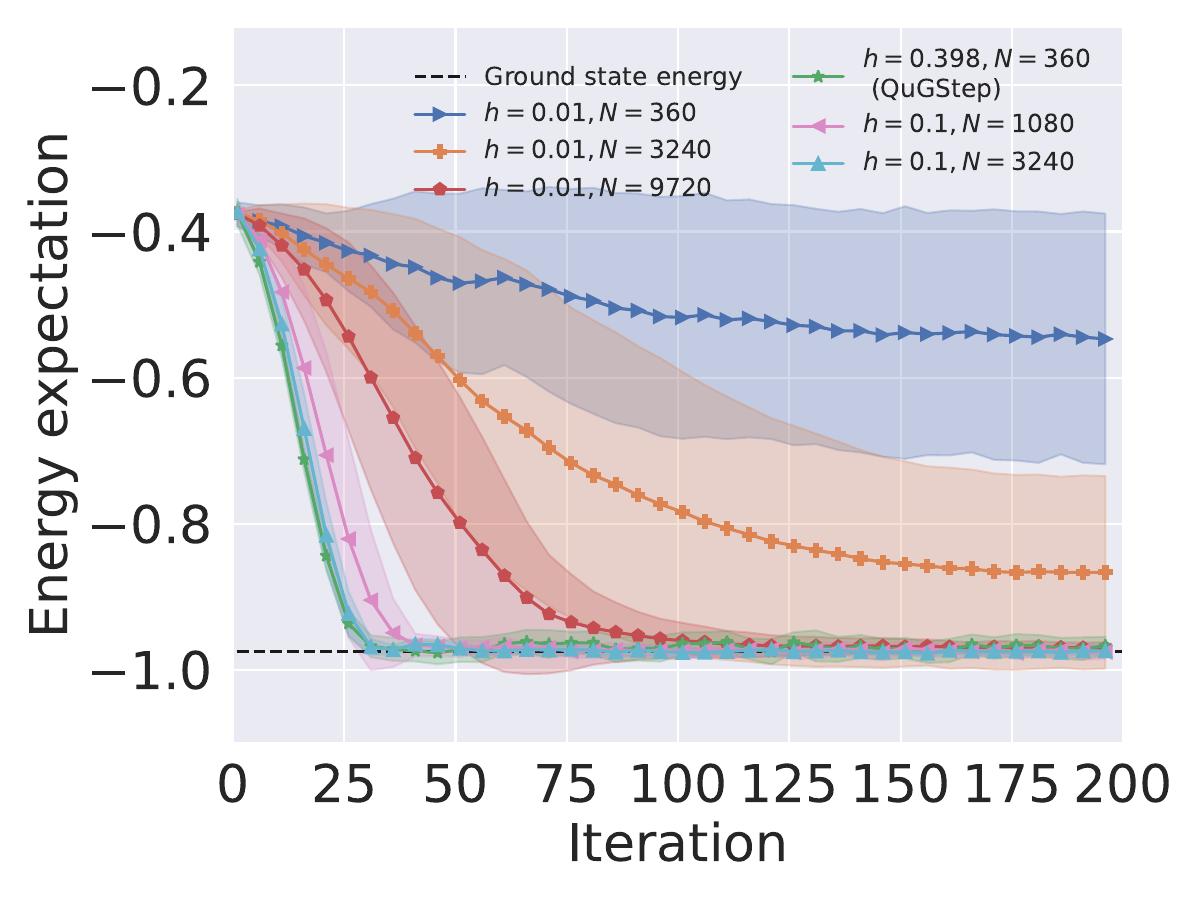}}
\caption{Optimizing the 1-parameter wavefunction in VQE to approximate the ground state energy of the H$_2$ molecule. The energy (Hartree) is plotted as a function of the number of iterations using various step sizes and the number of shots. In \textbf{(a)}, a step size of $h=1$ yields the lowest energy averaged over the last 20 iterations when $\hat{N}=9$, so we select $h_{9}\approx 1$ and accordingly $h_{360}$ can be selected as $h_9/(360/9)^{1/4}\approx 0.398$ using Eqn.~\eqref{eqn:scale}. \textbf{(b)} Comparison of energy performance profile of step size $h_{360}=0.398$ with those of other step sizes for $N=360$. \textbf{(c)} Comparison of energy performance profiles of varied shot budgets with fixed $h=0.01$, and QuGStep for $N=360$. The results in (a) are based on 5 independent experiments, while those in (b) and (c) are based on 30 experiments, with the solid curve representing the mean and the shaded area representing the standard deviation.}
 \label{fig:h2}
\end{figure*}

\begin{figure*}[ht]
\centering
	\subfigure[Test budget $\hat{N}=9$]{\includegraphics[width=0.33\linewidth ]{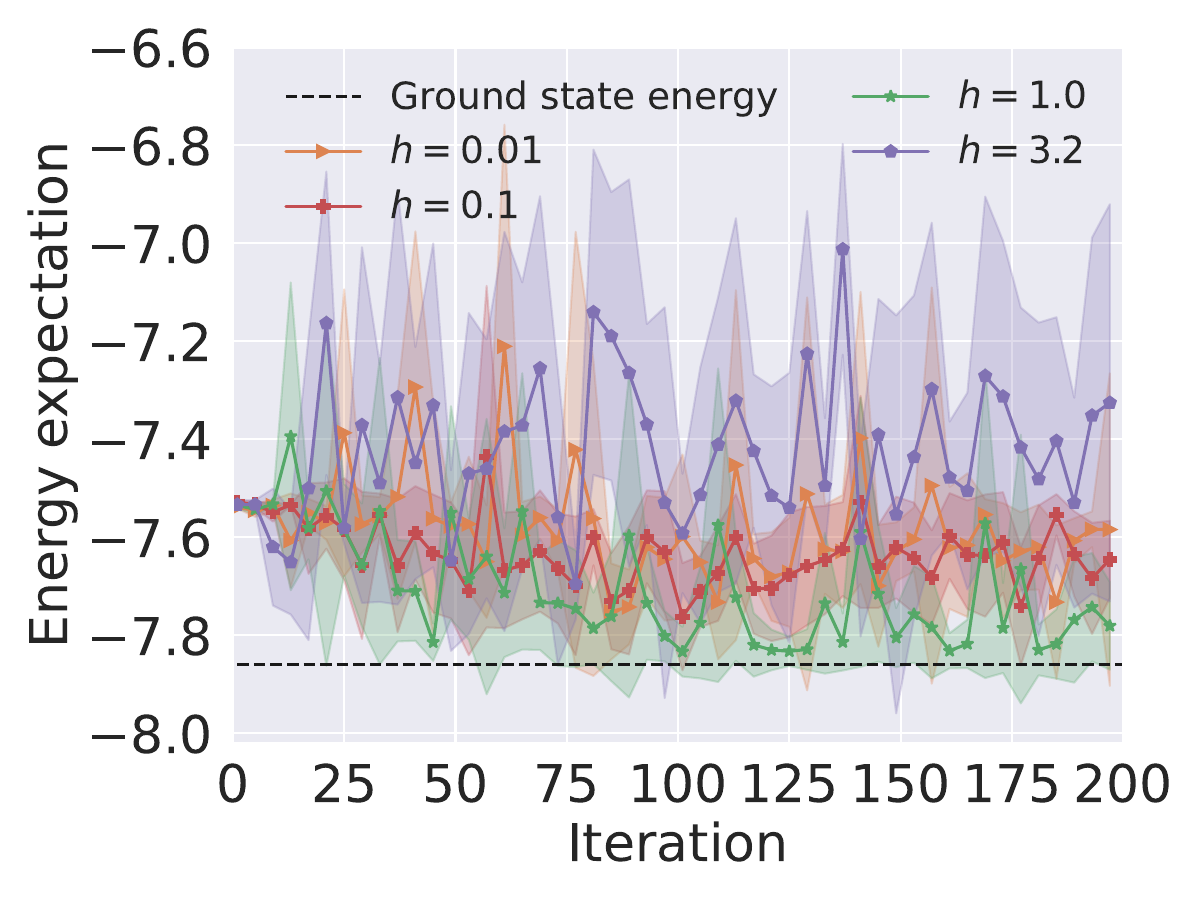}}
	\subfigure[Target budget $N=1800$]{\label{LiH1800}\includegraphics[width=0.33\linewidth ]{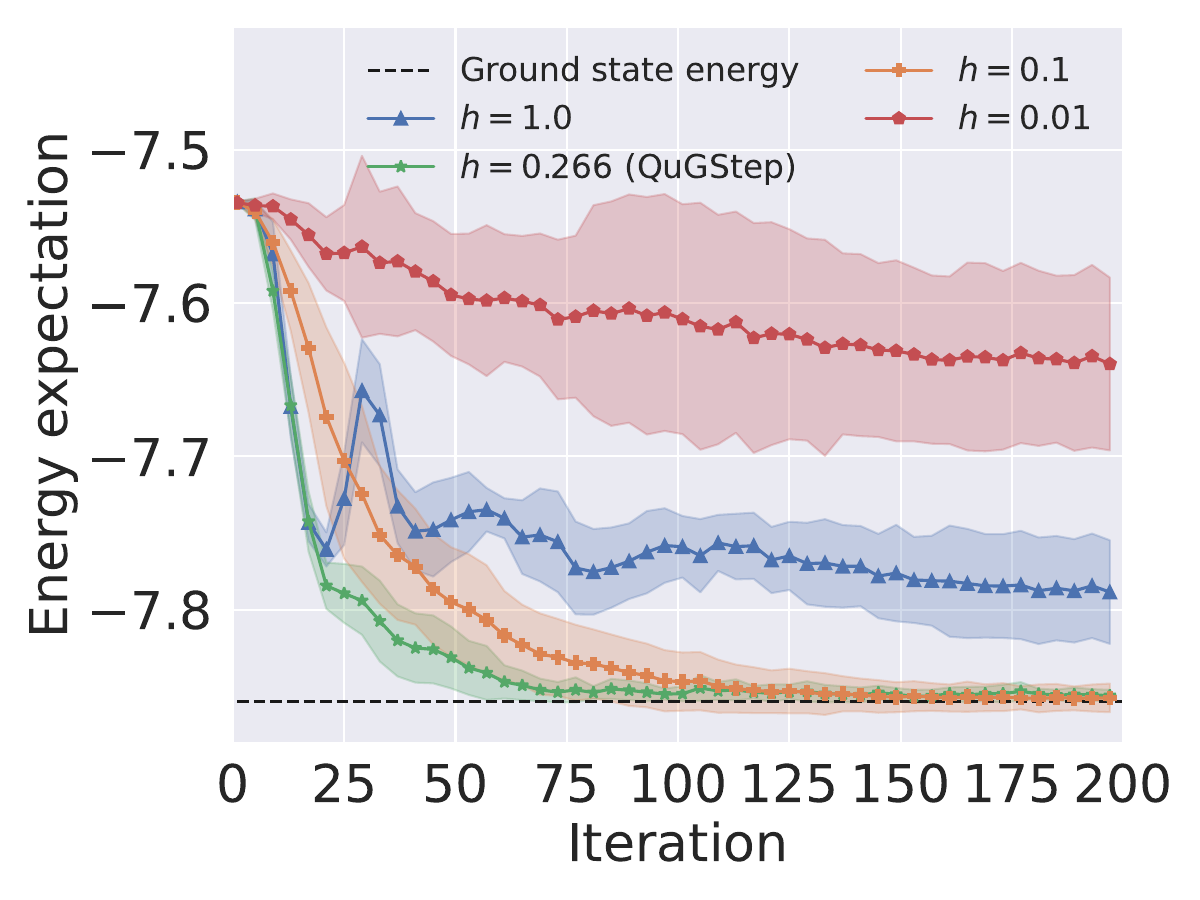}}
 \subfigure[Varied $N$]{\includegraphics[width=0.33\linewidth ]{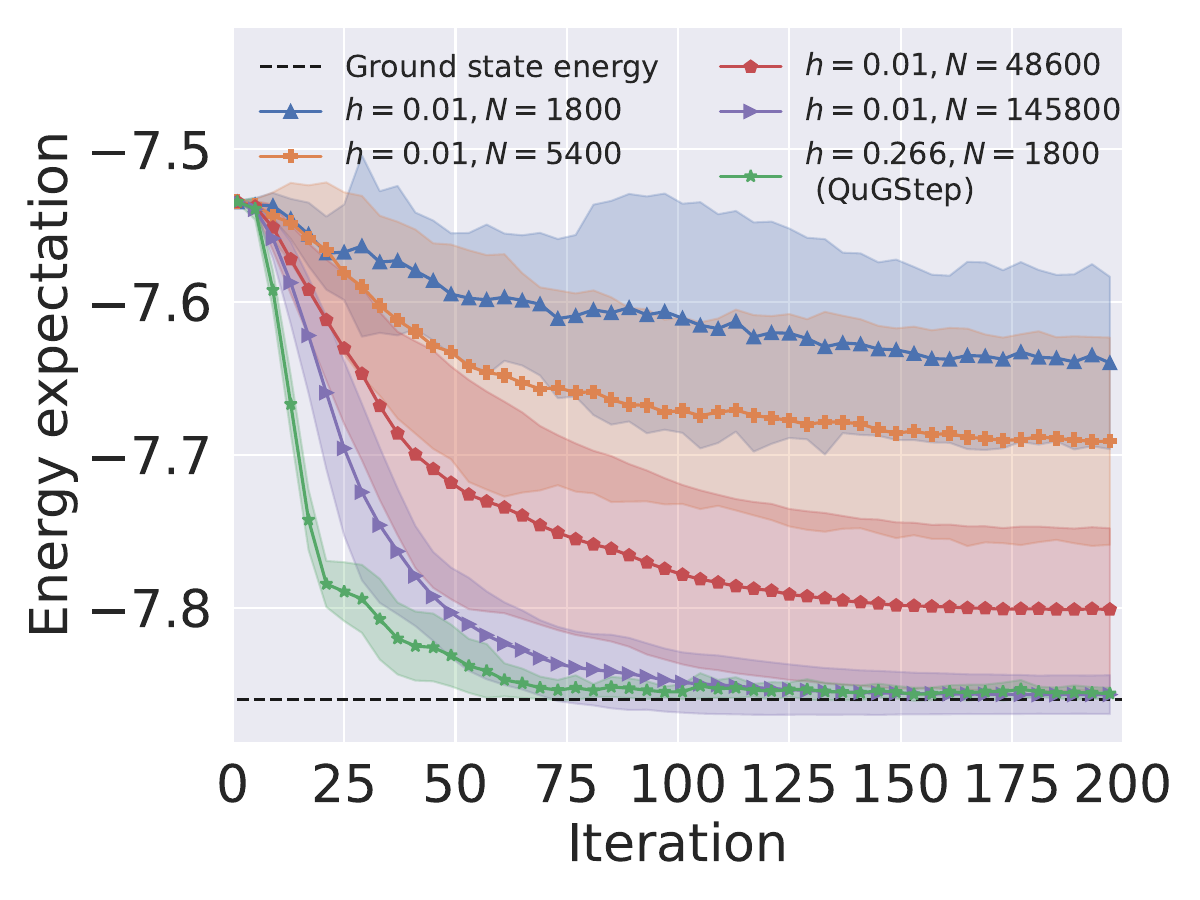}}
\caption{Optimizing the 8-parameter wavefunction in VQE to approximate the ground state energy of the LiH molecule. The energy (Hartree) is plotted as a function of the number of iterations using various step sizes and the number of shots. In \textbf{(a)}, although the energy curves for all step sizes oscillate significantly, a step size of $h=1$ yields the lowest energy averaged over the last 20 iterations when $\hat{N}=9$, so we select $h_{9}\approx 1$ and accordingly $h_{1800}$ can be selected as $h_{9}/(1800/9)^{1/4}\approx 0.266$ using Eqn.~\eqref{eqn:scale}. \textbf{(b)} Comparison of energy performance profile of step size $h_{1800}=0.266$ with those of other step sizes for $N=1800$. \textbf{(c)} Comparison of energy performance profiles of varied shot budgets with fixed $h=0.01$, and QuGStep for $N=1800$. The results in (a) are based on 5 independent experiments, while those in (b) and (c) are based on 20 experiments, with the solid curve representing the mean and the shaded area representing the standard deviation.}
 \label{fig:lih}
\end{figure*}

\begin{figure*}[ht]
\centering
	\subfigure[Test budget $\hat{N}=600$]{\includegraphics[width=0.33\linewidth ]{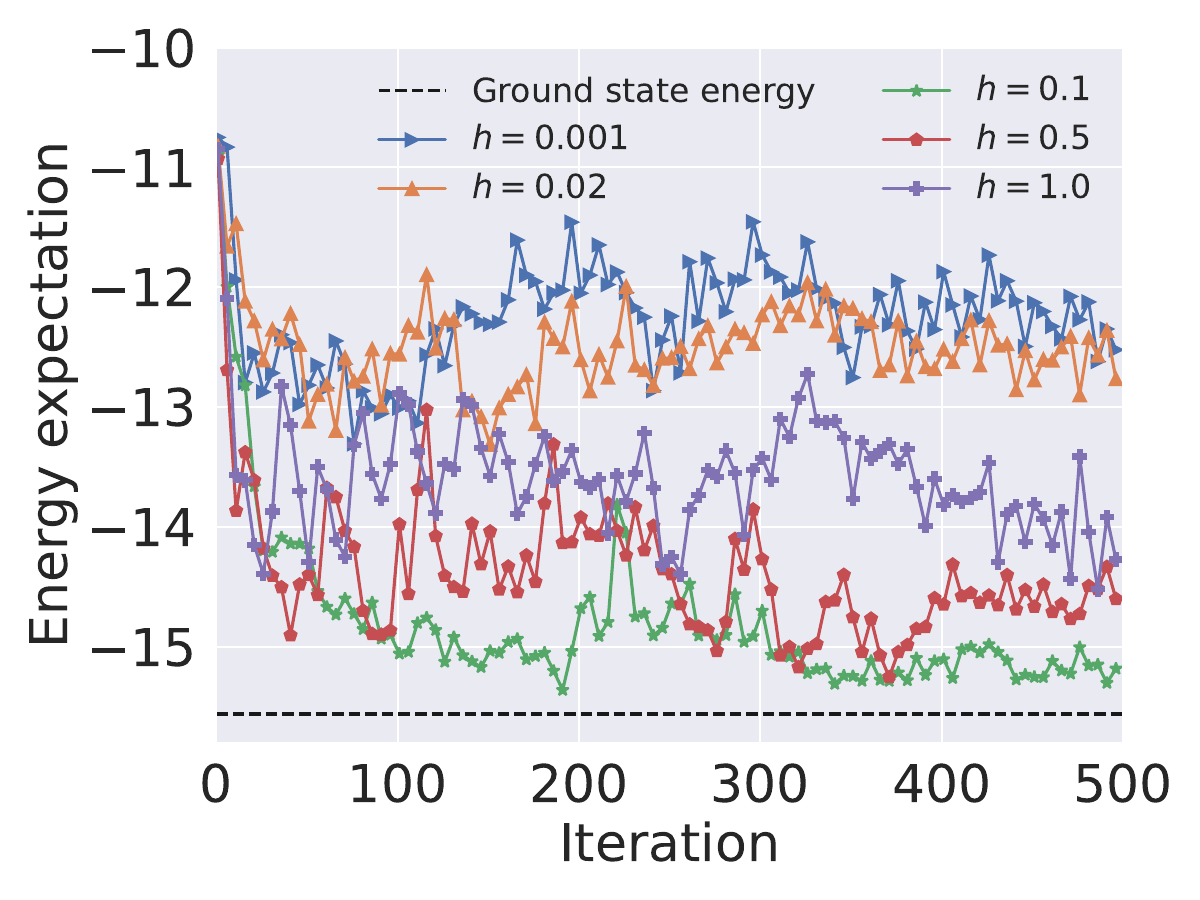}}
	\subfigure[Target budget $N=6000$]{\label{BeH2_6000}\includegraphics[width=0.33\linewidth ]{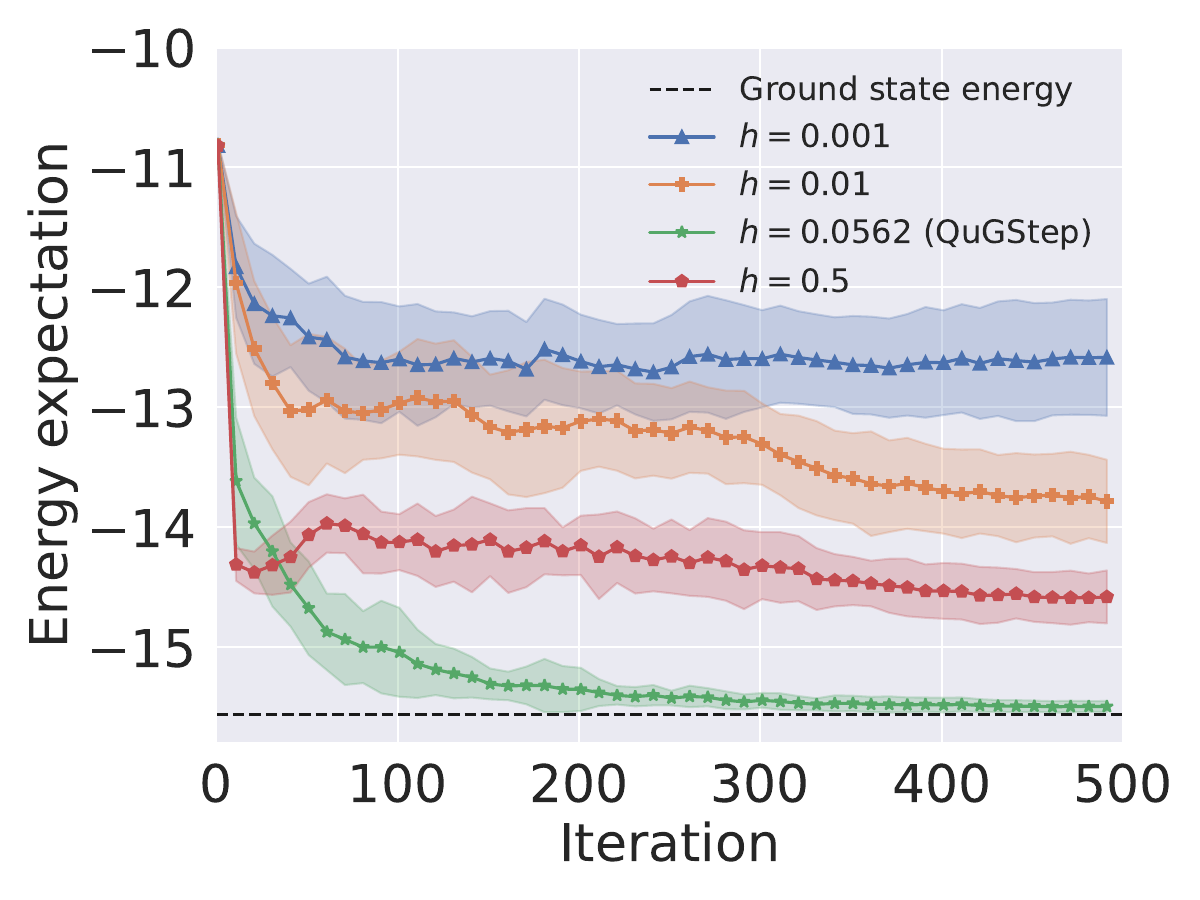}}
 \subfigure[Varied $N$]{\includegraphics[width=0.33\linewidth ]{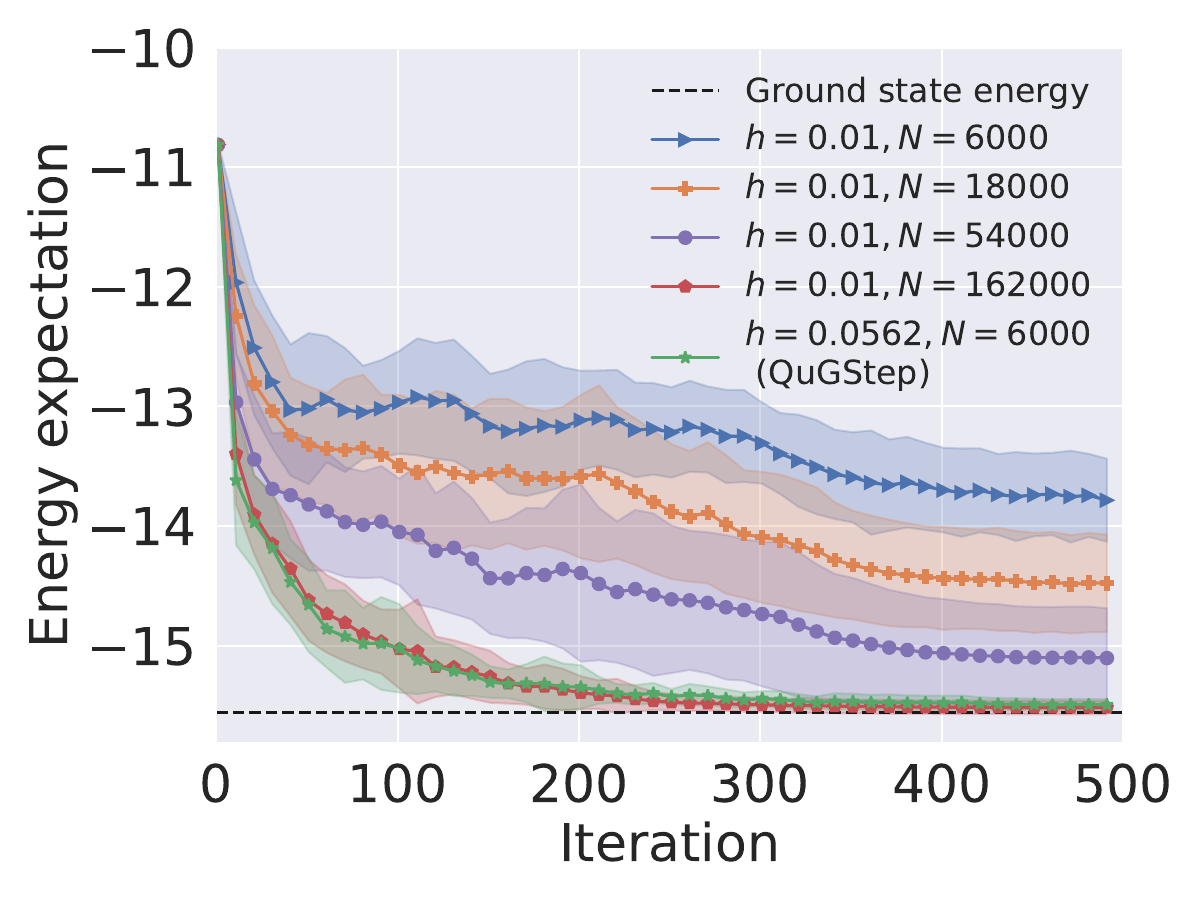}}
\caption{Optimizing the 36-parameter wavefunction in VQE to approximate the ground state energy of the BeH$_2$ molecule. The energy (Hartree) is plotted as a function of the number of iterations using various step sizes and the number of shots. In \textbf{(a)}, a step size of $h=0.1$ yields the lowest energy averaged over the last 20 iteration when $\hat{N}=600$, so we select $h_{600}\approx 0.1$ and accordingly $h_{6000}$ can be selected as $h_{600}/(6000/600)^{1/4}\approx 0.0562$ using Eqn.~\eqref{eqn:scale}. \textbf{(b)} Comparison of energy performance profile of step size $h_{6000}=0.0562$ with those of other step sizes for $N=6000$. \textbf{(c)} Comparison of energy performance profiles of varied $N$ with fixed $h=0.01$, and QuGStep for $N=6000$. Each curve in (a) is based on 1 experiment, while those in (b) and (c) are based on 20 experiments, with the solid curve representing the mean and the shaded area representing the standard deviation.}
 \label{fig:beh}
\end{figure*}

\begin{figure*}[ht]
\centering
 \subfigure[Adagrad]{\includegraphics[width=0.24\linewidth ]{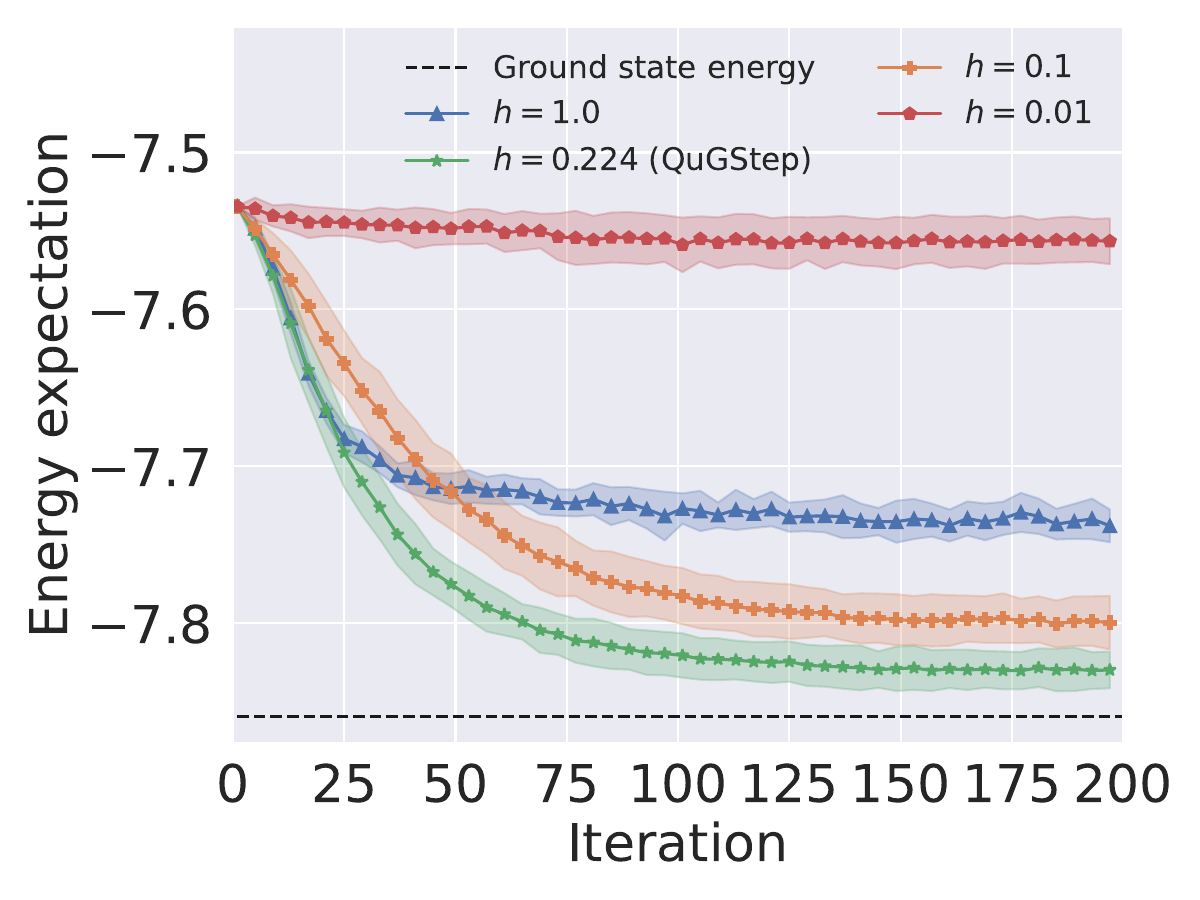}}
	\subfigure[MGD]{\includegraphics[width=0.24\linewidth ]{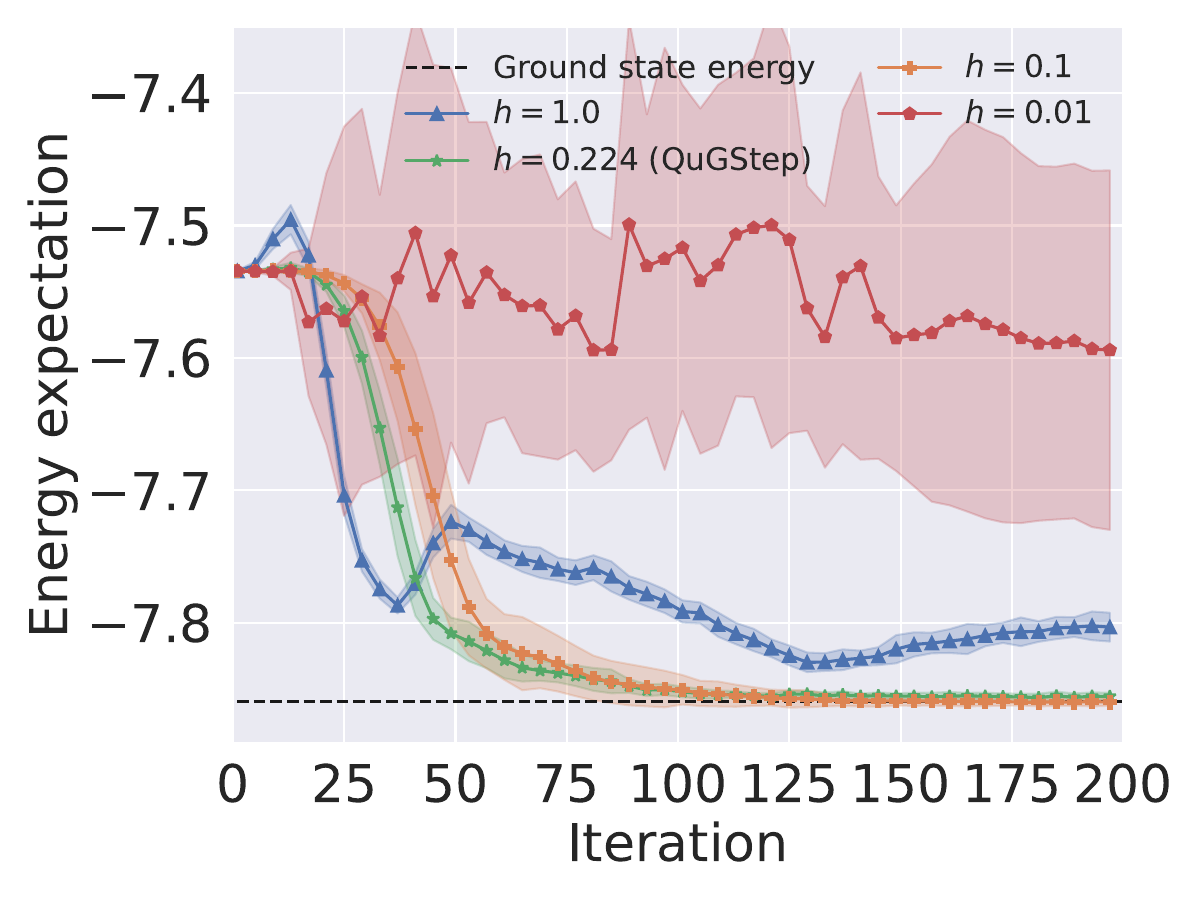}}
	\subfigure[RMSprop]{\includegraphics[width=0.24\linewidth ]{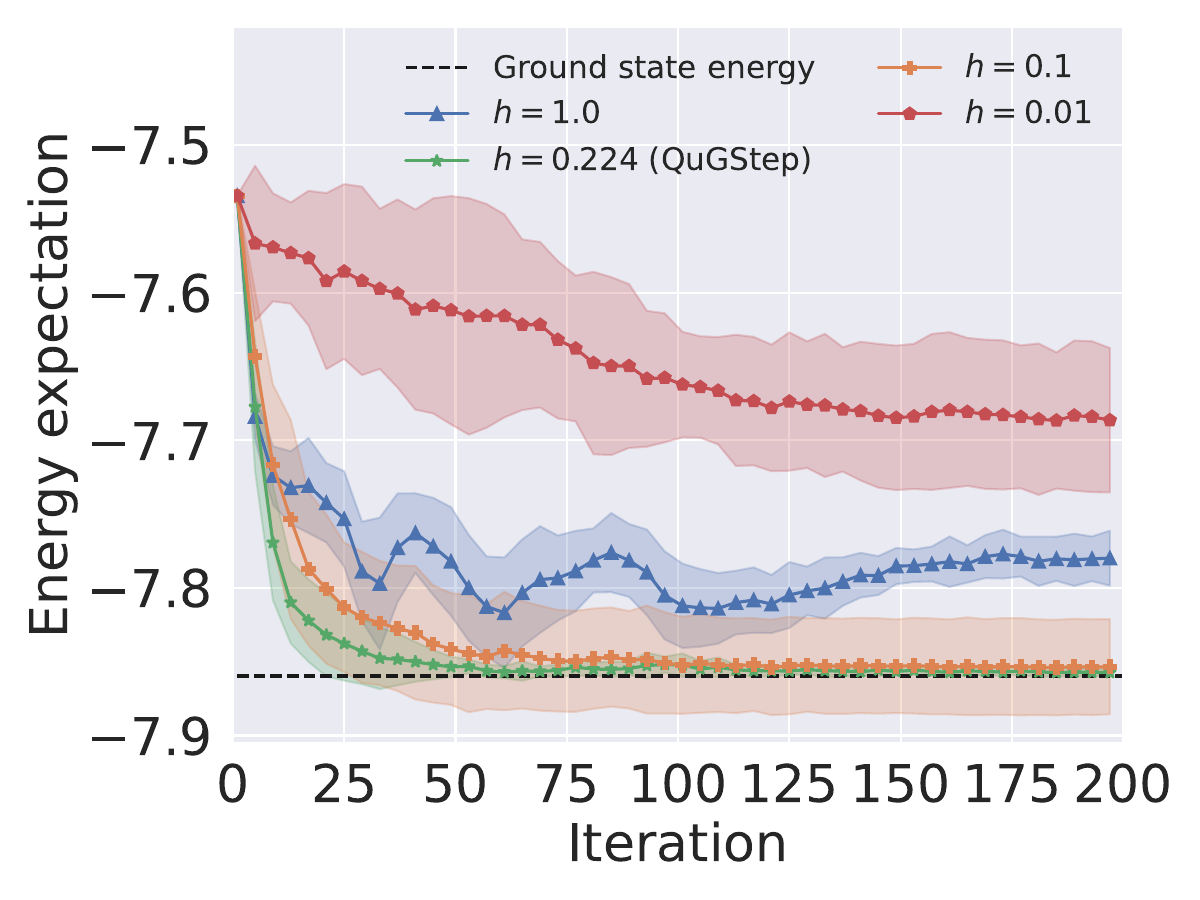}}
 \subfigure[Adam]
 {\includegraphics[width=0.24\linewidth ]{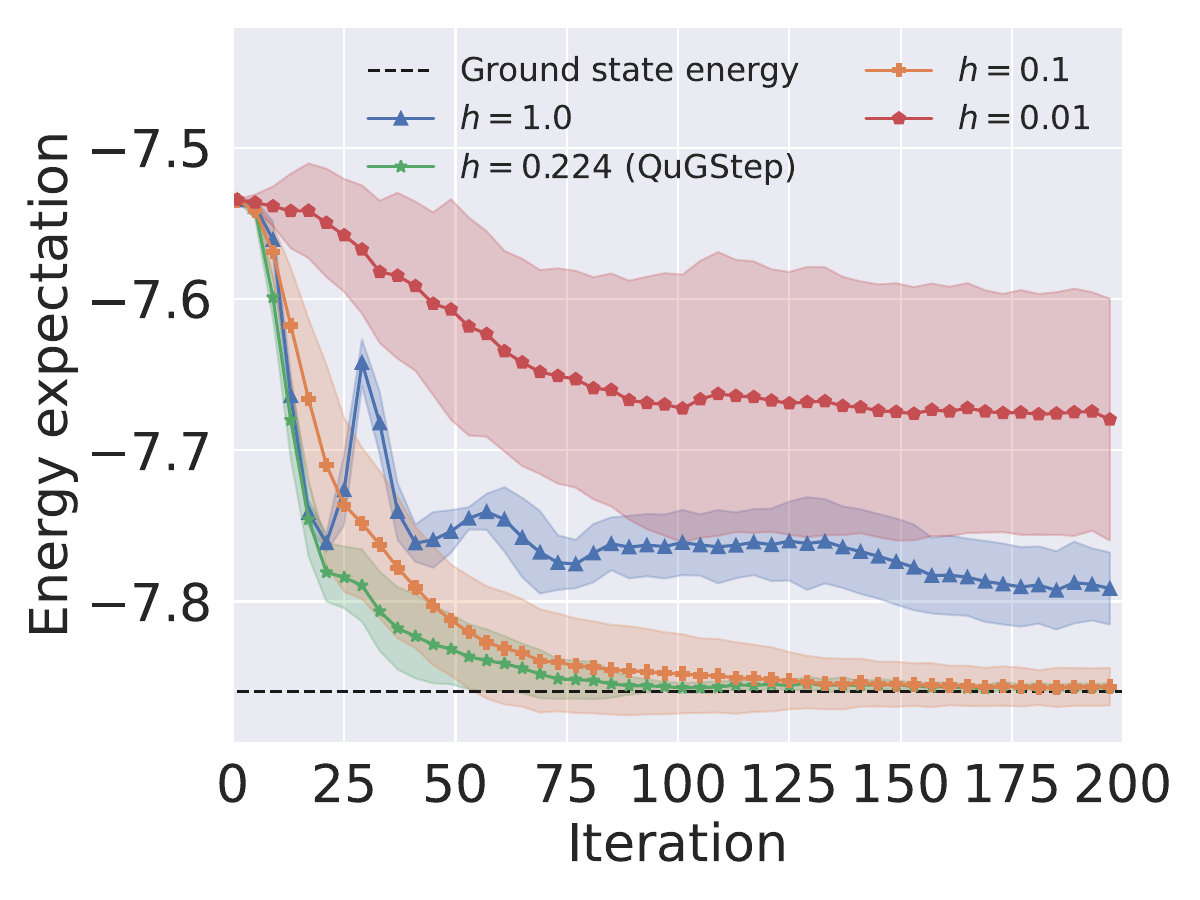}}
\caption{Different optimizers \textbf{(a)} Adagrad \textbf{(b)} momentum gradient descent (MGD) \textbf{(c)} RMSprop and \textbf{(d)} Adam for optimizing the 8-parameter wavefunction in VQE to approximate the ground state energy of the LiH molecule for $N=3600$. Since we obtained $h_9 \approx 1$ from Section \ref{sec:exp1}, $h_{3600}$ can be selected as $h_9/(3600/9)^{1/4} \approx 0.224$.}
 \label{fig:optimizer}
\end{figure*}

\section{Numerical experiments}
\label{sec:numerical}

In this section, we demonstrate the effectiveness of QuGStep for VQA optimization by a few numerical examples. To assess whether a finite difference step size is chosen appropriately, we examine whether such a choice affects the convergence of VQA. The experiments are primarily conducted to verify that:


\begin{itemize}
    \item The step size identified by QuGStep provides a more reliable gradient estimate for VQA than some commonly chosen default step sizes. To validate this, we compare the convergence of VQA optimization in which gradient estimations are obtained from finite difference approximations with different step sizes for a fixed measurement budget $N$. We then increase $N$ to check how large $N$ needs to set for the finite-difference step size of 0.01, a common choice in VQA~\cite{Guerreschi17_arXiv,Zhou20_021067}, to provide a reliable gradient estimate that ensure the convergence of VQA. While we use a step size of 0.01 in the comparison, we note that the default step size may vary across problems and is typically determined empirically.
    
    \item The step size identified by QuGStep indeed provides reliable and effective gradient estimates. To verify that, for the same problem, we use QuGStep step size to compute finite difference gradient estimates in several gradient based optimizers to check how the approximate gradient affect the convergence of these optimization algorithms.
    

\end{itemize}

\subsection{Experiment Setup}
We perform VQE to approximate the ground state energy of several molecular systems, including H$_2$, LiH and BeH$_2$. The experiments are conducted with the Qiskit QasmSimulator\cite{Qiskit}. Unless otherwise specified, the wavefunction parameters are optimized using the Adam optimizer~\ref{alg:grid}. The learning rate is initialized to 0.1 and follows a cosine decay strategy, where at iteration $t$ (out of $T$ total iterations), it is scaled by $0.5( \cos({\pi t}/{T}) + 1)$. For H$_2$ and LiH, the total number of iterations $T$ is set to 200, and for BeH$_2$, $T$ is set to 500. 

\textbf{H$_2$.} The H$_2$ molecule was modeled with an $H$-$H$ bond length set to $R_\mathrm{H_2} = 1.75$\AA~. For the quantum simulation, we employed the Unitary Coupled-Cluster Singles and Doubles (UCCSD) ansatz~\cite{Barkoutsos18_022322, O16_031007} combined with the the minimal STO-6G basis set. The molecular Hamiltonian is mapped to qubits using the Bravyi-Kitaev transformation, which results in a two-qubit system for the H$_2$ (see Appendix~\ref{app:h2}). The unitary transformation $U(\vec\theta)$ is defined by a single parameter $\theta$, expressed as: $U(\theta) = \text{exp}(-i\theta X_{0}X_{1})$. The simulation was initialized with the Hartree-Fock reference state  $\ket{\psi_{\text{ref}}} = \ket{\text{01}}$, which corresponds to the ground-state electron configuration in the molecular spin-orbital representation~\cite{O16_031007,Zhu24_2390}. This parameterization allows exploration of the ground-state energy landscape through variation of $\theta$, capturing essential electron correlation effects within the simplified representation.

\textbf{LiH.} In a minimal basis setup, the wave function of LiH can be represented using 12 qubits, where each qubit corresponds to a molecular spin orbital. By implementing the Parity transformation along with symmetries and excluding nonbonding orbitals, the Hamiltonian can be effectively simplified to a form that requires only four qubits~\cite{Rattew19_arXiv, Lolur23_789, Choy23_1206}. The transformation results in a Hamiltonian composed of 27 Pauli strings organized into 9 commuting cliques. The computation was performed at a bond length of $R_\mathrm{LiH} = 1.45$ \AA~. The minimal STO-3G basis set is used to represent the wavefunction. For efficient implementation on Noisy Intermediate-Scale Quantum (NISQ) devices, we employed a two-layer hardware-efficient circuit ansatz~\cite{Choy23_1197, Zhu24_2390} with 8 parameters. These parameters were initialized to zeros.


\textbf{BeH$_2$.} Starting from the STO-3G fermionic basis set, we reduced the Hamiltonian from a 10 spin-orbital basis to a six-qubit model via spin-parity mapping and qubit tapering at a molecular geometry of $R_\mathrm{BeH_2} = 1.5$\AA. We implemented a 3-layer hardware-efficient circuit with 36 parameters\cite{Kandala17_7671,Zhu24_2390}. The circuit parameters were initialized to 1.5. 

 
\subsection{QuGStep for Various Systems}
\label{sec:exp1}

We first apply QuGStep to determine the finite difference step size to be used for gradient estimations in the VQE optimizer applied to the H$_2$ molecule. The measurement budget for function evaluations is set to $N=360$. Specifically, we set $\hat{N}=9$ in Algorithm~\ref{alg:grid} and search for the best $h_{9}$ within $\mathds{S}=\{0.01, 0.1, 1, 10\}$. Figure~\ref{fig:h2}(a) shows the convergence history of different VQE runs associated with different choices of $h$. Because the computed energy expectation values obtained from VQE runs using $\hat{N}=9$ shots tends to be highly oscillatory, we conduct five runs for each candidate choice of $h_{9}$. Among these, $h_{9} = 1$ achieves the lowest average energy expectation value over the last 20 iterations. Consequently, we set $h_9 = 1$ and, based on scaling relation~\eqref{eqn:scale}, obtain $h_{360}\approx h_9/(360/9)^{1/4}\approx 0.398$. Figure~\ref{fig:h2} (b) shows that choosing $h_{360}=0.398$ enables VQE to converge to the lowest energy approximation. In contrast, an excessively large step size (e.g., $h = 1$) led to the convergence to a much higher local minimum, while a smaller step sizes, such as $h = 0.1$, slowed down convergence significantly. Choosing an even smaller step size (e.g., $h = 0.01$) led to convergence failure because the finite difference gradient estimation based on such a step size is completely dominated by noise. As shown in Figure \ref{fig:h2} (c), choosing $h=0.01$ as the finite difference step size would required a much more accurate function evaluated using a measurement budget of $9720$ shots to ensure VQE to converge to the ground-state energy. In contrast, QuGStep achieved convergence with only $N=360$ shots. After combining all the trial costs, we use a total of 540 shots ($9\times 4\times 5+360$) shots for each evaluation.  Compared to 9720 shots, the use of QuGstep results in approximately 94.4\% $(1-540/9720)$ reduction in measurement cost. Furthermore, even when using $h = 0.1$, a possibly more reasonable default in noisy settings, the required measurement budget ($N = 3240$) to achieve comparable performance to QuGStep’s $h \approx 0.398$ with $N = 360$ remains substantially higher. This still corresponds to an $\approx 83.3\%$ $(1 - 540/3240)$ reduction in measurement cost. Therefore, QuGStep may still yield significant savings relative to larger, more carefully and empirically chosen step sizes such as $h = 0.1$.

Next, QuGStep is applied to determine the finite difference step size to be used for gradient estimations in the VQE optimizer applied to the LiH molecule. The measurement  budget for function evaluations is \rv{set} to $N=1800$. We set $\hat{N}=9$ in Algorithm~\ref{alg:grid} and search for the best $h_9$ within $\mathds{S}=\{0.01, 0.1, 1, 3.2\}$.  Figure~\ref{fig:lih} (a) shows that $h=9$ achieves the lowest average energy expectation values. Therefore, we set $h_{9}=1$. It follows from \eqref{eqn:scale} that $h_{1800}\approx h_{9}/(1800/9)^{1/4}\approx 0.266$. Figure~\ref{fig:lih} (b) shows that choosing $h_{1800}= 0.266$ enables VQE to converge to the lowest energy approximation. Meanwhile, Figure \ref{fig:lih} (c) shows that choosing $h=0.01$ as the finite difference step size for gradient estimation requires us to increase the number of shot to 145800 in order to make VQE converge to the  ground-state energy. Taking into account all measurements used to choose an optimal $h_9$, each evaluation uses 1980 shots $(9×4×5 + 1800)$, which represents approximately 98.6\% $(1-1980/145800)$ reduction in measurement cost.

Finally, QuGStep is applied to determine the finite difference step size to be used for gradient estimations in the VQE optimizer applied to the BeH$_2$ molecule. The measurement budget for function evaluations is set to $N=6000$. We $\hat{N}=600$ in Algorithm~\ref{alg:grid} and search the best $h_{600}$ within the set $\mathds{S}=\{0.001, 0.02, 0.1, 0.5,1.0\}$. In Figure \ref{fig:beh} (a), we show that $h = 0.1$ achieves the lowest average energy expectation value. Therefore, we set $h_{600} = 0.1$. It follows from \eqref{eqn:scale} that $h_{6000}\approx h_{600}/(6000/600)^{1/4}\approx 0.0562$. As shown in Figure~\ref{fig:beh} (b), this choice of $h$ enables VQE to converge to the lowest energy approximation. Figure \ref{fig:beh} (c) shows that choosing a smaller step size 0.01 in the finite difference gradient estimation requires us to increase the number of shots to 162000 in order to obtain a sufficiently accurate gradient approximation to ensure the convergence of VQE to the ground-state energy of BeH$_2$. In contrast, using the step size identified by QuGStep for finite difference gradient approximation allow VQE to reach the same energy with only 6000 shots. This amounts to a 94.4\% reduction $(1-(600\times 5 + 6000)/162000)$ in the number of measurement shots.

As the system size increases, the number of parameters and the complexity of the energy landscape grow, typically requiring more optimization steps to reach the minimum. The results showed that approximate convergence could be achieved in about 30 iterations for H$_2$ (1 parameter) in Figure~\ref{H2_N360}, 75 iterations for LiH (8 parameters) in Figure~\ref{LiH1800}, and nearly the full 500 iterations for BeH$_2$ (36 parameters) in Figure~\ref{BeH2_6000}. In this study, we set $T$ to a fixed, large enough value for each molecule, which is a common practice when the precise point of convergence is unknown beforehand. 

\subsection{Effectiveness of $h_N$ Across Different Optimizers}

In this section, we aim to demonstrate that that the finite difference step size determined by QuGStep produces reliable and effective gradient estimates. To this end, we QuGStep to choose the finite different step size for gradient estimation in several optimization algorithms. These algorithms include the momentum gradient descent (MGD) algorithm \cite{Sutskever13_1139}, the RMSprop algorithm \cite{Graves13_arXiv} and the Adagrad algorithm \cite{Duchi11_7}.  We use the LiH system as the test problem. 

Following the discussion in Section~\ref{sec:exp1}, we set $\hat{N}=9$ and choose $h_9=1$.  For $N = 3600$, we obtain $h_{3600} = h_9/(3600/9)^{1/4} \approx 0.224$.  We use this  step size to perform finite difference gradient calculations in each optimization algorithm. As shown in Figure \ref{fig:optimizer}, choosing $h=0.224$ for finite difference gradient estimation works the best for all optimization algorithms. For Adagrad, none of the other tested step sizes allows the optimization algorithm to reach the ground-state energy in 200 iterations. Both MGD and RMSprop converge to the ground state with $h = 0.1$ and $h = 0.224$. However, choosing $h = 0.224$ for finite difference gradient estimation makes the optimization algorithm converge more quickly.

\section{Discussion}
\label{sec:discussion}
\textbf{The test budget and performance profile.} In principle, the test budget $\hat{N}$ should be significantly smaller than the target budget $N$. However, selecting the appropriate value for $\hat{N}$ requires careful consideration. If $\hat{N}$ is too small, multiple VQA runs may be necessary to stabilize the training curves, potentially increasing the overall cost of finding $h_{\hat{N}}$. Therefore, it is crucial to strike a balance between minimizing the test budget and ensuring reliable training outcomes.

Due to the noisy optimization process, particularly when using a small shot budget $\hat{N}$, the objective function curves may oscillate significantly. This variability complicates the assessment of the performance profiles. In this paper, we use the average values of the objective function from the last 20 iterations to evaluate the performance profile in the full run. We might also consider more efficient methods, such as applying a moving average to the objective function values or measuring the number of iterations required to achieve a certain threshold, or using an automated approach to determine these metrics~\cite{lin2024continuous,liang2025external}.

\textbf{Overhead of QuGStep.} The total overhead of QuGStep is approximately $|\mathds{S}|\times (\text{\# test runs})\times (\text{VQE iterations}) \times (d+1) \times \hat{N}$, where $|\mathds{S}|$ is the size of the candidate set. This indicates that the cost scales linearly with the number of parameters $d$. This observation suggests several potential strategies for future work, such as reducing the parameter subspace by using only a small, randomly selected subset of parameters to eliminate the $(d+1)$ factor, decreasing the number of VQE iterations in the test phase, since full convergence is unnecessary for distinguishing candidate step sizes, and replacing the simple grid search with more advanced methods, such as a midpoint search. 
 
\textbf{Noise type.} {The paper focuses solely on shot noise, excluding quantum noise caused by qubit and gate errors~\cite{wu2023qubits}, which cannot be mitigated by simply increasing the number of shots~\cite{Luo24_36}.  QuGStep is not directly applicable to such types of noise. In such scenarios, the error of gradient estimation will become $\mathcal{O}(h) + \mathcal{O}(1/(\sqrt{N}h))+ \mathcal{O}(1/h)$ instead of $\mathcal{O}(h) + \mathcal{O}(1/(\sqrt{N}h))$ as in Eqn.~\eqref{eqn:grad}, with the $\mathcal{O}(1/h)$ term arising from hardware noise. The optimal step size should balance the truncation error, which increases with $h$, against the noise terms, which decrease with $h$. Informally, we aim to balance $\mathcal{O}(h)$ with $\mathcal{O}(1/(\sqrt{N}h) + 1/h)$. Setting these two expressions equal, i.e., $\mathcal{O}(h) = \mathcal{O}(1/(\sqrt{N}h) + 1/h)$, we find that $h = \mathcal{O}(1/(N^{-1/2} + 1)^{1/2})$, establishing a relation between $h$ and $N$. This shows that $h$ still depends on $N$, though in a different way compared to the case with only shot noise. The exploration of incorporating other types of noise will be left for future work. } 

\textbf{Dynamical step size across different iterations.} Recent advances offer dynamic shot allocations across different optimization iterations~\cite{Zhu24_2390,Liang24_041403,Phalak23_41514}. In these dynamic shot allocation methods, it is also necessary to adjust the step size $h$ in accordance with changes in $N$ across iterations. We plan to explore this dynamic adjustment of the step size  $h$ for different iterations in future work.

\textbf{Extension to other gradient estimation method.} This paper considers the finite difference method because it is straightforward and generic approach
to gradient estimation and it only requires $d+1$ evaluations of the objective function for $d$ total parameters. The proposed methodology can be readily extended to other derivative estimation methods, such as central difference and the five-point stencil~\cite{abramowitz68_handbook}. However, these alternative methods generally require a larger number of objective function evaluations, which can lead to increased total shot noise.

\section{Conclusion}
\label{sec:conclusion}
We introduced an algorithm named QuGStep for choosing an optimal step size for finite difference gradient estimation when a fixed measurement budget is allocated for function evaluation in a VQA algorithm. QuGStep is developed by using the intuition that the optimal finite difference step size should be the one that minimizes an upper bound on the expected mean square error between the true gradient and its approximation.

Our numerical experiments, which involve using the VQE algorithm to compute the ground state energy of molecules such as H$_2$, LiH, and BeH$_2$, demonstrate that QuGStep effectively identifies an appropriate step size for gradient estimation for a fixed measurement budget. Using QuGStep, we achieve a significant reduction in the number of shots required in the VQE algorithm to converge to the ground state energy: 94\% for H$_2$
, 98\% for LiH, and 94\% for BeH$_2$, compared to VQE runs with the default step size  {\color{blue}(\textit{i.e.,} step size of $0.01$)}.

These results underscore QuGStep's ability to enhance the efficiency of VQA. This paper not only provides mathematical guidance for selecting the step size, but also emphasizes QuGStep's potential to improve the practical deployment and scalability of quantum computing technologies.
 
\begin{acknowledgments}
Y.C. and X.L. acknowledge the support to develop reduced scaling computational methods from the Scientific Discovery through Advanced Computing (SciDAC) program sponsored by the Offices of Advanced Scientific Computing Research (ASCR) and Basic Energy Sciences (BES) of the U.S. Department of Energy (\#DE-SC0022263).
This material is based upon work supported by the U.S. Department of Energy, Office of Science, Office of Advanced Scientific Computing Research and Office of Basic Energy Science, Scientific Discovery through Advanced Computing (SciDAC) program under Contract No. DE-AC02-05CH11231 and the Accelerated Research for Quantum Computing Program under Contract No. DE-AC02-05CH11231. Additional support is acknowledged from U. S. Department of Energy, Office of Science, National Quantum Information Science Research Centers, Quantum Systems Accelerator. This work used resources of the National Energy Research Scientific Computing Center (NERSC) using NERSC Award ASCR-ERCAP m1027 for 2023, which is supported by the Office of Science of the U.S. Department of Energy under Contract No. DE-AC02-05CH11231.
\end{acknowledgments}

\section*{Data Availability Statement}
The codes to reproduce the results presented in the paper are available on the GitHub repository:  \url{https://github.com/LeungSamWai/QuGStep}. These codes are released under the MIT license. 
\appendix

\section{Proof of Theorem 1}
\begin{proof} According to the Taylor expansion, we express $E(\tilde{\theta}+h)$ as follows:
$E(\tilde{\theta}+h)=E(\tilde{\theta})+E'(\tilde{\theta})h+\frac{1}{2}E''(\xi)h^2$, where $\xi \in (\tilde{\theta}-h, \tilde{\theta}+h).$ Consequently, we obtain 
\begin{align*}
&\mathcal{E}(\tilde{\theta}, h)=\left(\frac{\bar{E}\left(\tilde{\theta}+h\right)-\bar{E}\left(\tilde{\theta}\right)}{h}-E^{\prime}\left(\tilde{\theta}\right)\right)^2
\\=&(\frac{E(\tilde{\theta})+E'(\tilde{\theta})h+\frac{1}{2}E''(\xi)h^2+\epsilon_1-E(\tilde{\theta})-\epsilon_2}{h}-E'(\tilde{\theta}))^2
\\=&\left(\frac{1}{2}E''(\xi)h+\frac{\epsilon_1-\epsilon_2}{h}\right)^2
\\=&\frac{1}{4}|E''(\xi)|^2h^2+E''(\xi)(\epsilon_1-\epsilon_2)
+\frac{(\epsilon_1-\epsilon_2)^2}{h^2}.
\end{align*}
Here $\epsilon_1$ and $\epsilon_2$ are independent random variables with mean 0 and standard deviations $\frac{\sigma(\tilde{\theta}+h)}{\sqrt{N}}$ and $\frac{\sigma(\tilde{\theta})}{\sqrt{N}}$, respectively. Taking the expectation with respect to $\epsilon_1$ and $\epsilon_2$, we have 
\begin{align*}
&\mathds{E}\left(\frac{\bar{E}\left(\tilde{\theta}+h\right)-\bar{E}\left(\tilde{\theta}\right)}{h}-E^{\prime}\left(\tilde{\theta}\right)\right)^2
\\=&\frac{1}{4}|E''(\xi)|^2h^2+\frac{1}{h^2}\mathds{E}(\epsilon_1-\epsilon_2)^2
\\=&\frac{1}{4}|E''(\xi)|^2h^2+\frac{1}{h^2}\mathrm{Var}(\epsilon_1)+\frac{1}{h^2}\mathrm{Var}(\epsilon_2)\\ \leq& \frac{1}{4}\mu^2h^2+\frac{2\varsigma^2}{h^2N}, \forall h \in I.
\end{align*}
Applying the inequality $a + b \geq 2\sqrt{ab}$ for non-negative $a$ and $b$, we obtain $\frac{1}{4}\mu^2h^2+\frac{2\varsigma^2}{h^2N}\geq 2\sqrt{\frac{1}{4}\mu^2h^2\frac{2\varsigma^2}{h^2N}}$. Equality is achieved when $h$ is given by $h = \frac{8^{1/4}\varsigma^{1/2}}{\mu^{1/2}N^{1 / 4}}$.
\end{proof}

\section{H$_2$ Hamiltonian}\label{app:h2}
For the H$_2$ molecule, we follow the approach of O'Malley et al.~\cite{O16_031007}, using the Bravyi-Kitaev transformation to map the fermionic Hamiltonian to qubit operators. 
For H$_2$ in the minimal STO-6G basis at bond length $R = 1.75$ \AA, the molecular Hamiltonian after Bravyi-Kitaev transformation and exploiting symmetries reduces to a two-qubit system:
\begin{equation}
H = g_0 I + g_1 Z_0 + g_2 Z_1 + g_3 Z_0Z_1 + g_4 Y_0Y_1 + g_5 X_0X_1
\end{equation}

Following O'Malley et al., the UCCSD ansatz for this system simplifies to:
\begin{equation}
|\psi(\theta)\rangle = \exp(-i\theta X_0X_1)|\psi_{\text{ref}}\rangle
\end{equation}
where $|\psi_{\text{ref}}\rangle = |01\rangle$ is the Hartree-Fock reference state in the Bravyi-Kitaev basis, and the operator $X_0X_1$ generates transitions between the ground and excited electronic configurations.

\section{Leveraging Pauli Structure for Improved Error Bounds}\label{sec:errorbond}

When the parameterized quantum circuit consists of Pauli exponentials, we can derive more specific bounds for the second-order derivative in Theorem~\ref{thm}. Consider a single-parameter gate of the form:
\begin{equation}
U(\theta) = \exp(-i\theta P)
\end{equation}
where $P$ is a Pauli string with eigenvalues $\pm 1$. The parameterized state is $|\psi(\theta)\rangle = U(\theta)|\psi_{\text{ref}}\rangle$, and the energy expectation value is:
\begin{equation}
E(\theta) = \langle\psi_{\text{ref}}|U^\dagger(\theta)HU(\theta)|\psi_{\text{ref}}\rangle
\end{equation}

Using the Baker-Campbell-Hausdorff formula and the fact that $P^2 = I$ for Pauli operators, we can derive~\cite{Grimsley19_3007, Zhu22_033029}:
\begin{align}
\frac{\partial E}{\partial \theta} &= i\langle\psi(\theta)|[H,P]|\psi(\theta)\rangle \\
\frac{\partial^2 E}{\partial \theta^2} &= -\langle\psi(\theta)|[[P,H],P]|\psi(\theta)\rangle
\end{align}

Since $P$ is a Pauli string with norm $\|P\| = 1$, and using the operator norm inequality $\|[A,B]\| \leq 2\|A\|\|B\|$, we obtain:
\begin{equation}
\left|\frac{\partial^2 E}{\partial \theta^2}\right| \leq \|[[P,H],P]\| \leq 2\|[P,H]\| \leq 4\|H\|
\end{equation}

This provides a concrete bound $\mu \leq 4\|H\|$ for use in Theorem \ref{thm}. For molecular Hamiltonians in second quantization, $\|H\|$ can be bounded by:
\begin{equation}
\|H\| \leq \sum_{pqrs} |h_{pqrs}|
\end{equation}
where $h_{pqrs}$ are the molecular integrals.

However, this bound may be loose in practice. %

\nocite{*}
\bibliography{aipsamp.bib}

\end{document}